\pdfoutput=1
\documentclass[12pt]{article}
\usepackage{amsmath,amsfonts,amssymb,mathrsfs,graphicx}
\usepackage{epsfig}
\usepackage{cite}
\usepackage{multirow}
\usepackage{longtable}
\usepackage{rotating}
\usepackage{bbm}

\usepackage[squaren]{SIunits}

\allowdisplaybreaks[4] 

\newcommand{\capdef}{}
\newcommand{\mycaption}[2][\capdef]{\renewcommand{\capdef}{#2}%
        \caption[#1]{{\footnotesize #2}}}
\makeatletter
\renewcommand{\fnum@table}{\textbf{\tablename~\thetable}}
\renewcommand{\fnum@figure}{\textbf{\figurename~\thefigure}}
\makeatother

\newcounter{myenumi}

\renewcommand{\themyenumi}{\roman{myenumi}}
{\end{list}}

\newlength{\myem}
\newcounter{mysubequation}[equation]

\makeatletter
\renewcommand{\section}{\@startsection{section}{1}{0em}{-\baselineskip}%
{\baselineskip}{\normalfont\large\bfseries}}
\renewcommand{\subsection}%
{\@startsection{subsection}{2}{0em}{-0.7\baselineskip}%
{0.7\baselineskip}{\normalfont\bfseries}}
\makeatother

\newcommand{\ie}{{\it i.e.}}

\newcommand{\eg}{{\it e.g.}}

\newcommand{\cf}{{\it cf.}}

\newcommand{\eq}{Eq.}

\newcommand{\fig}{Fig.}

\newcommand{\Ref}{Ref.}
\newcommand{\Refs}{Refs.}

\newcommand{\equ}[1]{\eq~(\ref{equ:#1})}
\newcommand{\figu}[1]{\fig~\ref{fig:#1}}

\newcommand{\bi}{\begin{itemize}}
\newcommand{\ei}{\end{itemize}}

\begin{document}

\begin{titlepage}

\renewcommand{\thefootnote}{\alph{footnote}}

\vspace*{-3.cm}
\begin{flushright}
\end{flushright}


\renewcommand{\thefootnote}{\fnsymbol{footnote}}
\setcounter{footnote}{-1}

{\begin{center}
{\large\bf Constraints on the interpretation of the superluminal motion of neutrinos 
at OPERA
} \end{center}}
\renewcommand{\thefootnote}{\alph{footnote}}

\vspace*{.8cm}
\vspace*{.3cm}
{\begin{center} 
		{\large{\sc
                Walter~Winter\footnote[1]{\makebox[1.cm]{Email:}
                winter@physik.uni-wuerzburg.de}
                }}
\end{center}}
\vspace*{0cm}
{\it
\begin{center}

\footnotemark[1]%
       Institut f{\"u}r Theoretische Physik und Astrophysik, Universit{\"a}t W{\"u}rzburg, \\
       D-97074 W{\"u}rzburg, Germany

\end{center}}

\vspace*{1.5cm}

{\Large \bf
\begin{center} Abstract \end{center}  }

Various approaches aim to describe the recent analysis by the OPERA experiment, which indicates that neutrinos travel faster than the speed of light. We demonstrate that any such theoretical or experimental explanation must not destroy the complicated (nonlinear) structure of the proton waveform recovered in the neutrino signal. As one example, consider that only a fraction of the neutrinos travel faster than the speed of light, such as sterile neutrinos. We fit the OPERA data including this fraction as a free variable, assuming that the OPERA result is correct. In our analysis, the best-fit values are 50\% of the neutrinos being superluminal and $(v-c)/c = 4.5 \cdot 10^{-5}$, where the neutrino velocity increases as the fraction of superluminal neutrinos decreases.  The minimal fraction of superluminal neutrinos is found to be 17\% ($3 \sigma$), which is constrained by the non-linearity of the proton waveform. This minimal fraction challenges the hypothesis that only sterile neutrinos travel faster than the speed of light. In addition, we demonstrate that an experimental effect introducing a smearing between the proton waveform and neutrino signal, as expected for some systematical errors, is also limited by the shape of the waveform. Finally, we illustrate that even stronger constraints may be obtained from the recent analysis with a short-bunch beam, in spite of the low statistics.

\vspace*{.5cm}

\end{titlepage}

\newpage

\renewcommand{\thefootnote}{\arabic{footnote}}
\setcounter{footnote}{0}

\section{Introduction} 

Recently, the OPERA collaboration has measured the muon neutrino velocity, and (preliminarily) reported a difference between neutrino velocity and the speed of light  $(v-c)/c = 2.37 \cdot 10^{-5}$~\cite{Adam:2011zb}, which is compatible with an earlier MINOS result~\cite{Adamson:2007zzb}.  As far as the interpretation of this result is concerned, see \Refs~\cite{Ciborowski:2011ka,Alicki:2011qw,Drago:2011ua,Koch:2011qu,Matone:2011jd,Cohen:2011hx} for general comments, \Refs~\cite{Tamburini:2011ic,Klinkhamer:2011mf,Klinkhamer:2011iz,Mann:2011rd,Bi:2011nd,Lingli:2011yn,Alexandre:2011bu,Giudice:2011mm} for theoretical ideas indicating Lorentz invariance violation and \Ref~\cite{Ellis:2008fc} for earlier constraints, \Refs~\cite{Dvali:2011mn,Kehagias:2011cb,Wang:2011sz,Wang:2011oc,Saridakis:2011eq} for ideas involving environment-dependent effects or new couplings, \Refs~\cite{Autiero:2011hh,Anacleto:2011bv,Fargion:2011hd,Gardner:2011hg,GonzalezMestres:2011jc,Ciuffoli:2011ji,Konoplya:2011ag,Pfeifer:2011ve} for cosmological and astrophysical implications, and \Ref~\cite{Contaldi:2011zm,Knobloch:2011de} for the discussion of the analysis itself. In particular, any theory describing the superluminal motion has to face constraints such the bremsstrahlung issue discussed in \Ref~\cite{Cohen:2011hx}. We do not give any theoretical explanation of the superluminal motion, in particular, we do not answer the question if neutrinos can travel superluminally or not. Instead, we discuss the additional implications any theoretical or experimental description has to face, assuming that the result cannot be described by a trivial systematical error, such as an additional time correction not accounted for. More specifically, we show that the information  contained in the non-linear shape of the proton waveform is important, where we give a theoretical and an experimental example. 

As one of the major observations, the OPERA results are obviously inconsistent with very stringent limits from SN1987A~\cite{Hirata:1987hu,Bionta:1987qt,Longo:1987ub}. This implies either  a relatively strong energy dependence of the effect causing the superluminal motion of the neutrinos, see, \eg, \Refs~\cite{AmelinoCamelia:2011dx,Cacciapaglia:2011ax}, or a flavor- or mass-eigenstate dependence, see \Refs~\cite{Li:2011ue,Drago:2011ua,Mann:2011rd,Magueijo:2011xy,Cohen:2011hx,Giudice:2011mm} for related discussions. A particular mass-eigenstate dependent case are sterile neutrinos, such as taking a shortcut through an extra dimension~\cite{Pas:2005rb,Dent:2007rk,Hollenberg:2009ws,Nicolaidis:2011eq,Hannestad:2011bj}, see also \Ref~\cite{Arefeva:2011zp} for a similar approach.  Especially, sterile neutrinos may be an elegant way to circumvent the argument in \Ref~\cite{Cohen:2011hx}, since they do not couple to the electroweak gauge bosons.
On the other hand, the observation of neutrino oscillations is, in general, incompatible with the hypothesis that only some of the active states travel faster than the speed of light~\cite{Coleman:1997xq,Coleman:1998ti}. However, this argument does not necessarily apply if only the sterile neutrinos are affected, since the incoherent limit, where active-sterile oscillations are averaged out, is perfectly consistent with short-baseline oscillation fits. In addition, note that neutrino oscillations have not been observed in OPERA yet ($\nu_\tau$ appearance can be described by flavor mixing), which means that even the propagation of the active neutrinos may be incoherent.  In this letter, we
use the most general assumption that only a fraction $X$ of the observed neutrinos are superluminal. This fraction can be  related to one or more of the active or sterile mass eigenstates, as it may be plausible for a tachyonic neutrino.  In addition, we test the hypothesis if the signal may due to some additional smearing not accounted for in the analysis, which may be coming from some unknown systematics. Finally, we comment on the recent test with a short-bunch wide-spacing beam.

\section{Neutrino propagation}

The OPERA experiment is sensitive to $\nu_\mu \rightarrow \nu_\mu$, where the fraction of the mass eigenstate $\nu_i$ in $\nu_\mu$ is $|U_{\mu i}|^2$. For the best-fit values from a recent global fit~\cite{Schwetz:2011zk}, $|U_{\mu 1}|^2 \simeq 0.14$, $|U_{\mu 2}|^2 \simeq 0.35$, and $|U_{\mu 3}|^2 \simeq 0.51$ if only active neutrinos exist, which means that mostly $\nu_3$ are produced in the OPERA beam. Since only $\nu_\mu$ are detected, the transition probability contains another factor of $|U_{\mu i}|^2$. For instance, if the propagation of the neutrinos were incoherent, one would have
\begin{equation}
 P_{\mu \mu} = \sum\limits_i | U_{\mu i} |^4 = \sum\limits_{i=1}^3 | U_{\mu i} |^4 + \sum\limits_{s, \, \text{sterile}} | U_{\mu s} |^4 \label{equ:pmm}
\end{equation}
in the presence of active and sterile states. For the current best-fit values, $|U_{\mu 1}|^4 \simeq 0.02$, $|U_{\mu 2}|^4 \simeq 0.12$, and $|U_{\mu 3}|^4 \simeq 0.26$, which means that $\nu_3$ clearly dominates. 

We assume that the fraction $X$ of the muon neutrinos detected in OPERA is superluminal, traveling with a common speed $(v-c)/c$, and the fraction $1-X$ is subluminal, traveling roughly with light speed. This assumption is very general in the sense that it does not depend on coherence (and, therefore, neutrino oscillations), which may only apply to a subset of the states  -- such as the active ones. For the special case of incoherent propagation, the fraction $X$ of superluminal neutrinos can be written as (\cf, \equ{pmm}) 
\begin{equation}
 X \equiv \frac{\sum\limits_{i, \, \text{superluminal}} | U_{\mu i} |^4}{\sum\limits_{i, \,  \text{subluminal}} | U_{\mu i} |^4+\sum\limits_{i, \, \text{superluminal}} | U_{\mu i} |^4} \, , \label{equ:defx}
\end{equation}
where, in general, all active and sterile states are summed over. If there are only active states, $X \simeq 0.05$ for (only) $\nu_1$ superluminal, $X \simeq 0.30$ for  (only) $\nu_2$ superluminal, and $X \simeq 0.65$ for  (only) $\nu_3$ superluminal. If there are sterile states, for instance, a current short-baseline analysis of neutrinos oscillations~\cite{Kopp:2011qd} favors, in a 3+2 model, $|U_{\mu 4}|=0.165$ and $|U_{\mu 5}| = 0.148$ at the best-fit (and also close to the incoherent limit, which may be applicable here), which means that  $|U_{\mu 4}|^4+|U_{\mu 5}|^4 \simeq 0.0012$. Thus, if only the sterile states are superluminal, one has $X \simeq 0.003$ in this simple model. In general, however, larger values in the percent range are potentially not excluded, whereas values $X \gg 0.1$ are inconsistent with other long-baseline disappearance data, see, \eg, \Ref~\cite{Adamson:2011ku}.

\section{Method}

\begin{figure}[t]
\begin{center}
\includegraphics[width=10cm]{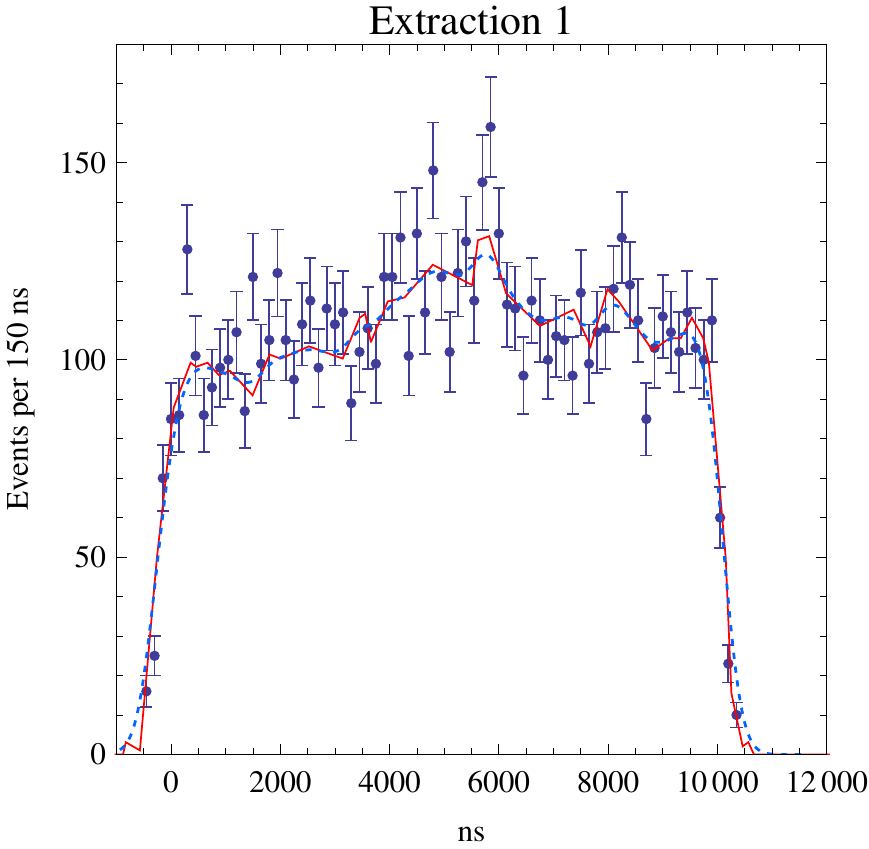}
\end{center}
\mycaption{\label{fig:compopera}Comparison of the measured neutrino interaction time distribution (data points) and the best-fit of the proton waveform (shifted by $1043.4 \, \mathrm{ns}$; solid curve); figure reproduced from preliminary result in \Ref~\cite{Adam:2011zb} (v2). The dotted curve shows the proton waveform with a hypothetical (Gaussian) filter with a width of 200~ns applied. }
\end{figure}

The measurement of $(v-c)/c$ is based on the time delay $\delta t$ of the actual arrival time of the neutrinos compared to the expected arrival time assuming speed of light propagation, where a number of known corrections are taken into account and subtracted in the original analysis. For the derivation of this time delay, the shape of the proton waveform, measured by a beam current transformer, is matched to the shape of the neutrino signal.
We have first reproduced the result ($\delta t=57.8$~ns)  from \Ref~\cite{Adam:2011zb}, where we have only considered the statistical error and their first extraction. We show in \figu{compopera} the corresponding comparison of the measured neutrino interaction time distribution (data points) and the proton waveform function (solid curve, shifted by $1043.4 \, \mathrm{ns}$, which includes a known 985.6~ns correction), reproduced from \Ref~\cite{Adam:2011zb} within the error from the resolution of the figures. Our analysis uses a Gaussian $\chi^2$ with a free normalization of the proton waveform to be fitted as another parameter. However, $\delta t$ and the proton waveform normalization have turned out to be uncorrelated, which means that the normalization has been determined and fixed for the further analysis. One can easily see from \figu{compopera} that not only the rising and falling edges between proton waveform and neutrino signal match very well, but also the nonlinear structure at the plateau. Thus, any effect modifying this structure will lead to a reduced goodness of fit.

Our minimal $\chi^2$ per d.o.f. is slightly larger (1.3) than the one of the OPERA analysis (1.1). We find that $\delta t=0$ can be excluded at $5 \sigma$, roughly consistent with the OPERA result, and our statistical error is slightly larger (about 11~ns, compared to 8~ns in the original analysis). These differences are expected from our rough approximation of the proton waveform, and from the different statistical method used here.
We have also tested the impact of using more bins for the rising and falling edges of the the measured neutrino interaction time, without qualitative change. Since we prefer a self-consistent picture, we base our analysis on the waveform in \figu{compopera}. Note that we do not consider systematical uncertainties quantified by the OPERA collaboration here, since they only add to the statistical uncertainty of the time of flight measurement in a trivial way, and cannot affect the shape of the waveform to the degree we discuss -- as we will illustrate below.

\section{Fraction of superluminal neutrinos}

\begin{figure}[t!]
\begin{center}
\includegraphics[width=10cm]{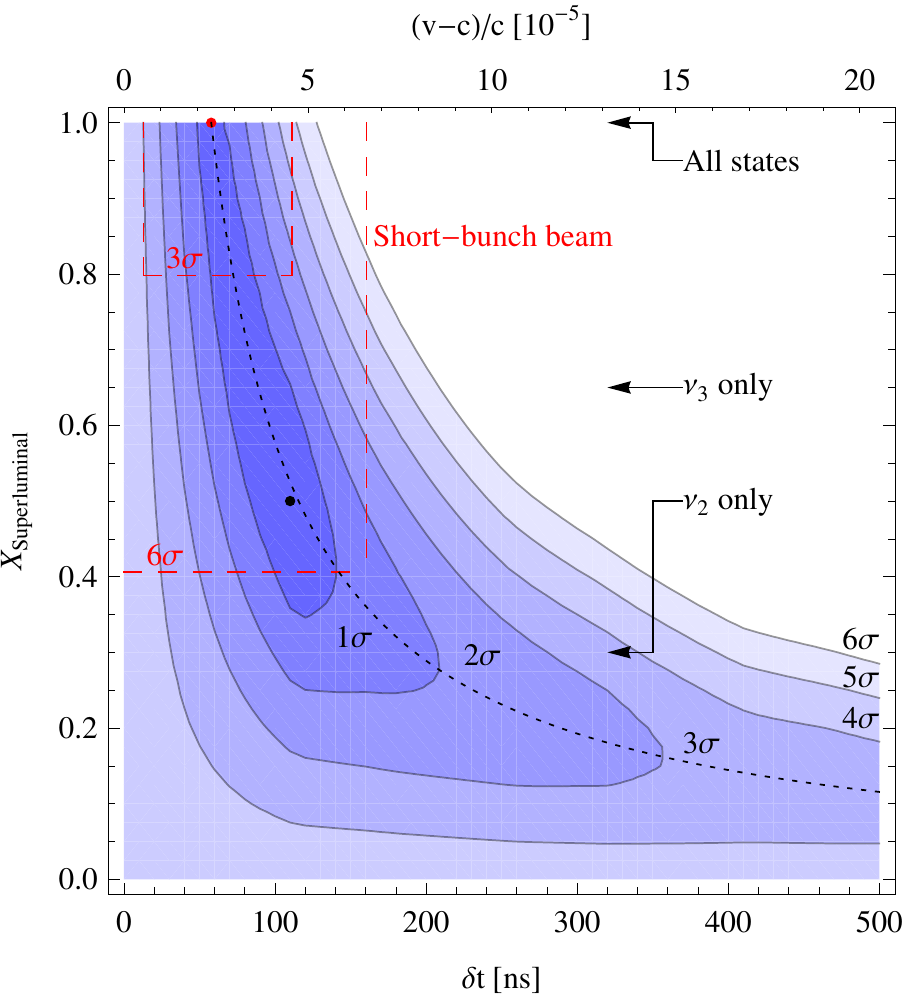}
\end{center}
\mycaption{\label{fig:fracx} Combined fit of $\delta t$ (or $(v-c)/c$, on upper horizontal axis) and $X$ from the OPERA result (2 d.o.f.). Here only the statistical error is included. The fractions $X$ expected from \equ{defx} for specific superluminal mass eigenstates (All states, $\nu_3$ only, $\nu_2$ only) in the incoherent limit are also shown. The (preliminary) OPERA result corresponds to $X\equiv 1$, with the best-fit marked by the red (upper) dot. The two-parameter best-fit is marked by the black (lower) dot. An additional constraint, which may be obtained from the test with a short-bunch wide-spacing beam, is shown by the dashed regions. The dotted curve represents the parameter degeneracy in \equ{deg}.}
\end{figure}

We assume that a fraction $X$ of the detected neutrinos travels superluminally, and a fraction $1-X$ subluminally. We have first of all observed that a smaller $X$ can be easily compensated for by a larger $\delta t$. This near-degeneracy comes from the approximate linearity of the proton waveform at the rising and falling edges (\cf, \figu{compopera}), which dominate the measurement. In the linear regime at the edges, one can show that the degeneracy 
\begin{equation}
X \times \delta t \simeq  57.8 \, \mathrm{ns}
\label{equ:deg}
\end{equation} 
holds. However, a too small value of $X$ will lead to a change of the spectral shape in the non-linear regime, such as on top of the proton waveform, which limits the fraction $X$ from below.

We show our main result in \figu{fracx} as a  combined fit of $\delta t$  and $X$ from the OPERA result (2 d.o.f.). The OPERA result corresponds to $X\equiv 1$, with the best-fit marked by the red (upper) dot. It corresponds to the case of all states superluminal. If $X<1$, one can clearly see the degeneracy with $\delta t$, \ie, larger $\delta t$ can compensate for smaller $X$ (\cf, dotted curve). In fact, the $5\sigma$  contour extends to very large values of $\delta t$. This means that if $X$ is marginalized over, $\delta t$ can be hardly constrained from the above. However, it can be clearly seen that $\delta t=0$ can be excluded for any $X$ at about $5 \sigma$, including the special case $\delta t=X=0$. The best-fit is found at $\delta t= 110$~ns and $X=0.5$, where the $\chi^2$ is relatively flat for $X \gtrsim 0.4$. Therefore, only a lower bound for $X$ can be determined, which is $X \gtrsim 0.17$ at $3 \sigma$ (1 d.o.f., since projected on vertical axis).

As far as the interpretation of our result is concerned, let us focus on \equ{defx} first, which assumes the incoherent limit. It can be read off in this simplistic approach that  the OPERA result is consistent with only $\nu_2$ or $\nu_3$ being superluminal (see arrow markers in figure), whereas only $\nu_1$ can be excluded. Of course, also combinations of the states or even all three states could be superluminal. If only sterile neutrinos are superluminal, such as by shortcuts though an extra dimension, the corresponding fraction $X$ has to be $X \gtrsim 0.17$ in order to be compatible with the results at $3 \sigma$. Note that in this case, the active states may still oscillate. Such a large contribution $X$ of sterile neutrinos is, however, implausible with respect to short baseline data for sterile neutrinos in the electronvolt range. Therefore, theories with superluminal sterile neutrinos can probably be ruled out -- unless $X$ in the detector is especially enhanced for the OPERA experiment parameters, and \equ{defx} is severely modified; see, \eg, \Ref~\cite{Bramante:2011uu} for a specific model, which demonstrates how to use this information in specific models. 

In the most recent OPERA paper~\cite{Adam:2011zb} (v2), a test with a short-bunch wide-spacing beam was performed in order to trace back the individual neutrinos to specific proton bunches. In this test, no neutrinos with $\delta t=0$ have been observed. This imposes an even stronger constraint on $X$, in spite of the low statistics of only 20~events. If, for instance, the distribution of $\delta t$ values is divided into two bins (split at $\delta t = 31~\mathrm{ns}$), at least 80\% of the neutrinos have to be superluminal at $3 \sigma$ from a simple Possonian $\chi^2$-test (including mis-identification of events, which can be estimated from the rms). In addition, $\delta t$ is directly constraint from the distribution of $\delta t$-values, where the error can be obtained from the measured rms. In \figu{fracx}, these new constraints are shown as dashed regions for $3 \sigma$ and $6\sigma$. However, these bounds based on the assumption of a Gaussian distribution of the shape of the signal ($\delta t$ values), which is hard to confirm for the given low statistics. More general, it cannot be uniquely established that there is only one velocity component.

\section{Gaussian averaging of proton waveform}

\begin{figure}[t]
\begin{center}
\includegraphics[width=10cm]{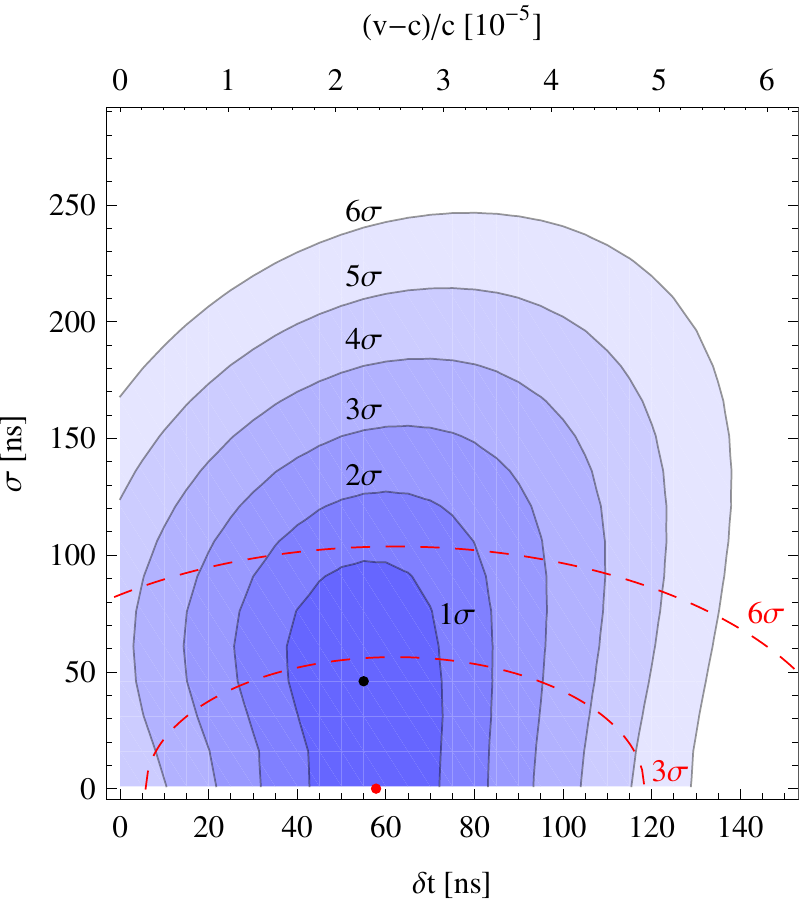}
\end{center}
\mycaption{\label{fig:filter} Combined fit of $\delta t$ (or $(v-c)/c$, on upper horizontal axis) and a filter width $\sigma$ from the OPERA result (2 d.o.f.). Here only the statistical error is included.  The (preliminary) OPERA proton waveform corresponds to $\sigma \rightarrow 0$, with the best-fit marked by the red (lower) dot. The two-parameter best-fit is marked by the black (upper) dot. An additional constraint, which may be obtained from the test with a short-bunch wide-spacing beam, is shown by the dashed regions. }
\end{figure}

In \Ref~\cite{Knobloch:2011de}, it was speculated that an additional smearing effect, not accounted for in the original analysis, may cause a deformation of the rising and falling edges of the neutrino waveform compared to the proton waveform, or even be implied in the proton waveform if not too large. Likewise, many systematical errors may not only cause a systematical time shift, such as the ones discussed by the OPERA collaboration, but also introduce some smearing. This includes also theoretical effects, such as non-negligible extensions of the neutrino wavepackets, see, \eg, \Ref~\cite{Naumov:2011mm} (Fig.~4). On the other hand, the shape of the proton waveform is convincincly well recovered in the neutrino signal, \cf, \figu{compopera}, even in the nonlinear regime. Therefore, the impact of such systematical errors cannot be too large. In order to quantify this statement for this particular  class of systematical errors, we assume that such a smearing could be Gaussian in the simplest case, around a mean offset in $\delta t$. For example, the dotted curve in \figu{compopera} shows a smeared proton waveform  where Gaussian smearing with a width of 200~ns has been used. Obviously, this smearing deforms the rising and falling edges. However, it also deforms the spectral features on top of the proton waveform, which will eventually reduce the goodness of fit.  We leave the width $\sigma$ of such a Gaussian smearing as a free variable and fit it together with $\delta t$ in \figu{filter}. As it can be read off from the figure, the best-fit $\delta t$ hardly changes in the presence of a possible smearing. Around the value $\sigma \simeq 50$~ns, where the rising and falling edges are deformed, the confidence level for the exclusion of $\delta t=0$ slightly decreases. However, from the top of the contours, $\sigma \lesssim 140 \, \mathrm{ns}$ at the $3 \sigma$ CL (1 d.o.f.), where a significant contribution to the $\chi^2$ comes from the nonlinear part of the proton waveform. This implies that even asymmetric effects must be limited if they introduce an smearing over such timescales.
In addition, from \figu{filter}, $\delta t=0$ can be excluded at the $4 \sigma$ confidence level, even if $\sigma$ is marginalized over, which means that the OPERA result cannot have been compromised by such a large smearing effect (if Gaussian). In the meanwhile, as expected, no such effect has been found  by the test with the short-bunch beam. In fact, if the shape of the distribution of events is assumed to be Gaussian, an additional constraint can be derived, shown as dashed curves in \figu{filter}. In this case, the width of the smearing $\sigma$ is stronger limited. However, given the low statistics, the Gaussian nature of the distribution cannot yet be unambiguously established.

Note that the class of systematics discussed here may also impact the measurement of $X$ discussed above, if present, since the measurement of $X$ relies on the shape of the proton waveform. On the other hand, one can learn from \figu{filter} that smearing effects at the magnitude of the systematical errors discussed by the OPERA collaboration (up to 8.3~ns) cannot distort the shape of the proton waveform significantly (in addition to introducing a systematical shift). This may be different for the short-bunch measurement, where additional systematical errors dominate the measurement.

\section{Summary and conclusions}

We have demonstrated that theoretical or experimental interpretations of the OPERA result changing the shape between proton waveform and neutrino signal lead to a reduced goodness of fit. We have used two examples: the fraction of neutrinos being superluminal, and possible smearing effects between between proton waveform and neutrino signal.

Assuming that only a fraction $X$ of the neutrinos is superluminal, the best-fit is found to be at $X=0.5$ with $\delta t=110$~ns, corresponding to $(v-c)/c \simeq 4.5 \cdot 10^{-5}$. This result is in fact closer to the MINOS result than the original OPERA result. However, fractions  $X \gtrsim 0.35$ are equally plausible. Most importantly, $X$ can be limited from below $X \gtrsim 0.17$ ($3 \sigma$), which means that superluminal sterile neutrinos will be challenged. On the other hand, only $\nu_2$ or $\nu_3$ superluminal may be consistent with the OPERA data from the time of flight measurement. This result will be, however, not compatible with neutrino oscillations if neutrino oscillations can be established in OPERA. Even stronger constraints can be obtained from the test with a short-bunch beam in spite of the low statistics, where no neutrinos with the speed of light have been seen. Therefore, in conclusion, the most plausible scenario to describe the OPERA result, if correct, is that all eigenstates travel faster than the speed of light, whereas flavor- or mass-eigenstate approaches to reconcile the OPERA and MINOS with SN1987A and other data are not promising.

We have also demonstrated that there cannot be a significant smearing effect in the extrapolation from the proton to the neutrino waveform, as it may be expected for a class of systematical errors not accounted for. This result is supported by the short-bunch run, although the shape of the signal cannot yet be uniquely determined in this case.

\subsubsection*{Acknowledgments}

I would like to thank Marcos Dracos and Joachim Kopp for useful discussions, and 
I would like to acknowledge support from Deutsche Forschungsgemeinschaft, grants WI 2639/3-1 and WI 2639/4-1.


\end{document}